\documentclass[12pt]{article}
\usepackage{amsfonts}
\usepackage{epsfig,amssymb,euscript}
\usepackage{amsmath,amscd}

\numberwithin{equation}{section}
\begin{document}

\begin{titlepage}
\begin{center}
\renewcommand{\thefootnote}{\fnsymbol{footnote}}
{\Large{\bf Solutions of Minimal Four Dimensional de Sitter Supergravity}}
\vskip1cm
\vskip 1.3cm
J. B. Gutowski$^1$  and W. A. Sabra$^2$
\vskip 1cm
{\small{\it
$^1$Department of Mathematics\\
King's College London\\
Strand\\
London WC2R 2LS, UK\\}}
\vskip .6cm {\small{\it
$^2$ Centre for Advanced Mathematical Sciences and Physics Department\\
American University of Beirut\\ Lebanon  \\}}
\end{center}
\bigskip
\begin{center}
{\bf Abstract}
\end{center}

Pseudo-supersymmetric solutions of minimal $N=2$, $D=4$ de Sitter supergravity are classified 
using spinorial geometry techniques. 
We find three classes of solutions. The first class of solution consists of geometries which are
fibrations over a 3-dimensional manifold equipped with a Gauduchon-Tod structure.
The second class of solution is the cosmological Majumdar-Papapetrou solution 
of Kastor and Traschen, and the third corresponds to gravitational waves propagating in
the Nariai cosmology.

\end{titlepage}

\section{Introduction}

The classification of supersymmetric solutions has attracted considerable
attention in recent years due to the important role these solutions play in
string and M-theory. Many years ago, Tod was able to find all metrics
admitting supercovariantly constant spinors in $N=2$, $D=4$ ungauged minimal
supergravity \cite{tod}. In recent years and motivated by the work of 
\cite{tod},  progress has been made in the classification of supersymmetric
supergravity solutions. 
For example, the classification of
supersymmetric solutions of minimal $N=2$, $D=4$ gauged supergravity has
been performed in \cite{klemm2003, klemm2004}. The bosonic part of $N=2$, 
$D=4$ gauged supergravity is basically Einstein-Maxwell theory with a
negative cosmological constant. The supersymmetric solutions of this theory
are obtained by solving the Killing spinor equation obtained from the
vanishing of the gravitini supersymmetry variation. \ In this paper, we will
be interested in finding solutions of Einstein-Maxwell theory with a
positive cosmological constant. This theory cannot be embedded in a
supergravity theory, as supersymmetry restricts the cosmological constant to
be negative. \ However, one can obtain a fake Killing spinor equation by
analytic continuation.\ Therefore we will be using fake supersymmetry as a
solution generating technique. De Sitter supergravities can also be obtained from
type IIB* theory \cite{Liu:2003qaa}. Solutions of five dimensional De Sitter
supergravity were recently analyzed in \cite{hkt}.
We shall use spinorial geometry techniques to investigate solutions
of the minimal four-dimensional de Sitter supergravity.
These have been used to analyse supersymmetric solutions
in ten and eleven dimensional supergravity theories
\cite{papadgran2005a, papadgran2005b, papadgran2006a, papadgran2006b}
as well as in lower dimensions \cite{roest2007}.

The plan of the paper is as follows. In Section two, a summary of the basic
equations of the theory of $N=2$, $D=4$  minimal de Sitter supergravity and
a brief description of the representatives of the Killing spinors are
presented. Sections three, four and five contain a detailed analysis of the
Killing spinor equations for the three canonical forms of the Dirac spinors,
making use of the linear system presented in the Appendix. We conclude in
Section six.

\section{$N=2$, $D=4$ Minimal de Sitter Supergravity}

In this section, we present a summary of $N=2$, $D=4$ minimal de Sitter
supergravity. The bosonic action associated with this theory is 
\cite{freedman, fradkin} 
\begin{equation}
S=\int \ \mathrm{d}^{4}x  \sqrt{-g} \ \big({\frac{1}{4}}R-{\frac{1}{4}}F_{\mu \nu }F^{\mu
\nu }-{\frac{3}{2\ell ^{2}}}\big)
\end{equation}
where $F=dA$ is the Maxwell field strength and $\ell $ is a non-zero real
constant. The signature of the metric is $(-,+,+,+)$.
The Einstein and gauge field equations are 
\begin{eqnarray}
R_{\mu \nu} -3 \ell^{-2} g_{\mu \nu} -2 F_{\mu \rho} F_{\nu}{}^\rho +{1 \over 2} 
F_{\alpha \beta}F^{\alpha \beta} g_{\mu \nu} &=&0
\nonumber \\
 \hskip 6cm d \star F &=& 0 \ .
\end{eqnarray}

We shall consider solutions which are \textit{pseudo-supersymmetric} i.e.
those which admit a non-zero Killing spinor $\epsilon$ satisfying the
Killing spinor equation: 
\begin{eqnarray}  \label{kse}
D_{\mu }\epsilon =\bigg( \partial _{\mu }+{\frac{1}{4}}\omega _{\mu ,\nu
_{1}\nu _{2}}\Gamma ^{\nu _{1}\nu _{2}}+{\frac{i}{4}}F_{\nu _{1}\nu
_{2}}\Gamma _{\mu }\Gamma ^{\nu _{1}\nu _{2}}
\nonumber \\ \hskip2cm
-iF_{\mu \nu }\Gamma^{\nu }
-{\frac{i}{2}}\ell ^{-1}\Gamma _{\mu }-\ell ^{-1}A_{\mu }\bigg) \epsilon =0 \ .
\end{eqnarray}
The Killing spinor $\epsilon$ is a Dirac spinor. We follow the conventions
of \cite{gutsabd4} in dealing with such spinors; for convenience we
summarize a number of useful results here.

Dirac spinors can be written as complexified forms on $\mathbb{R}^{2}$; a
generic spinor $\eta $ can therefore be written as \cite{lawson}

\begin{equation}
\eta =\lambda 1+\mu ^{i}e_{i}+\sigma e_{12}
\end{equation}
where $e_1$, $e_2$ are 1-forms on $\mathbb{R}^2$, and $i=1,2$; $e_{12}=e_1
\wedge e_2$. $\lambda $, $\mu ^{i}$ and $\sigma $ are complex functions. It
will be particularly useful to work in a null basis, and set 
\begin{eqnarray}
\Gamma _{+} &=&\sqrt{2} i_{e_{2}}  \nonumber \\
\Gamma _{-} &=&\sqrt{2} e_{2}\wedge  \nonumber \\
\Gamma _{1} &=&\sqrt{2} i_{e_{1}}  \nonumber \\
\Gamma _{\bar{1}} &=&\sqrt{2} e_{1}\wedge
\end{eqnarray}
where in the null basis the metric is 
\begin{eqnarray}
ds^2 = 2 \mathbf{e}^+ \mathbf{e}^- +2 \mathbf{e}^1 \mathbf{e}^{{\bar{1}}} \ .
\end{eqnarray}

We recall from \cite{gutsabd4} that a spinor $\epsilon$ can be written,
using $Spin(3,1)$ gauge transformations, as one of three possible simple
canonical forms: 
\begin{eqnarray}  \label{form1}
\epsilon = e_2
\end{eqnarray}
or 
\begin{eqnarray}  \label{form2}
\epsilon = 1+ \mu^1 e_1
\end{eqnarray}
or 
\begin{eqnarray}  \label{form3}
\epsilon = 1+ \mu^2 e_2 \ .
\end{eqnarray}
Note that by making use of a $Spin(3,1)$ transformation generated by 
$\Gamma_{+-}$, combined with an appropriately chosen $U(1)$ gauge
transformation of $A$ which together leave $1$ invariant, one can without
loss of generality take $|\mu^2|=1$ in ({\ref{form3}}).

To proceed, we evaluate the Killing spinor equation ({\ref{kse}}) acting on
the spinor 
\begin{eqnarray}
\epsilon = \lambda 1+\mu^{i}e_{i} \ .
\end{eqnarray}
The resulting equations are summarized in Appendix A. We then consider the
three cases ({\ref{form1}}), ({\ref{form2}}) and ({\ref{form3}}) separately.

\section{Solutions with $\protect\epsilon=e_2$}

In order to analyse solutions with $\epsilon =e_{2}$, evaluate the equations
in Appendix A with $\lambda =\mu ^{1}=0$ and $\mu ^{2}=1$. One obtains 
\begin{equation}
F_{+-}+F_{1{\bar{1}}}+\ell ^{-1}=0
\end{equation}
and 
\begin{equation}
F_{+-}+F_{1{\bar{1}}}-\ell ^{-1}=0 \ .
\end{equation}
It is clear that these equations admit no solution; hence there are no
supersymmetric solutions with $\epsilon =e_{2}$.

\section{Solutions with $\protect\epsilon =1+\protect\mu e_{1}$}

On evaluating the equations in Appendix A with $\lambda=1$, $\mu^1=\mu$, 
$\mu^2=0$ one one obtains the conditions:

\begin{eqnarray}  \label{muder}
\partial _{1}\mu &=&\partial _{+}\mu =0  \nonumber \\
\partial _{\bar{1}}\mu &=&{\sqrt{2}i}\ell ^{-1}\left( 1+\mu \bar{\mu}\right)
\end{eqnarray}
\begin{eqnarray}
\omega _{1,+1} &=&\omega _{-,+1}=\omega _{1,+{\bar{1}}}= 
\omega _{+,1{\bar{1}}}=\omega _{+,+{\bar{1}}}=0  \nonumber \\
\omega _{-,1{\bar{1}}} &=&\frac{\left( \mu \partial _{-}\bar{\mu}-\bar{\mu}
\partial _{-}\mu \right) }{\left( 1+\mu \bar{\mu}\right) }  \nonumber \\
\omega _{1,1{\bar{1}}} &=&-\mu {\sqrt{2}i}\ell ^{-1}
\end{eqnarray}
\begin{eqnarray}
\ell ^{-1}A_{-} &=&\frac{1}{2}\frac{\left( \bar{\mu}\partial _{-}\mu +
\mu\partial _{-} \bar{\mu} \right) }{\left( 1+\mu \bar{\mu}\right) } -{\frac{1}{2}}
\omega _{-,+-}  \nonumber \\
\ell ^{-1}A_{{1}} &=&-{\frac{1}{2}}\omega _{{1},+-} -{\frac{1}{\sqrt{2}}}\mu 
{i}\ell ^{-1}  \nonumber \\
\ell ^{-1}A_{+} &=&-{\frac{1}{2}}\omega _{+,+-}
\end{eqnarray}
\begin{eqnarray}
F_{-1} &=&-\frac{i}{\sqrt{2}}\frac{\partial _{-}\mu }{\left( 1+ \mu \bar{\mu}
\right) }  \nonumber \\
F_{+-} &=&\ell ^{-1}  \nonumber \\
F_{1{\bar{1}}} &=&F_{+1}=0 \ .
\end{eqnarray}

To analyse these conditions, observe first that 
\begin{eqnarray}
d \mathbf{e}^- = - \omega_{A,+-} \mathbf{e}^A \wedge \mathbf{e}^-
\end{eqnarray}
and hence $\mathbf{e}^-$ is hypersurface orthogonal; one can introduce a
co-ordinate $u$ and function $H$ such that 
\begin{eqnarray}
\mathbf{e}^- = H du \ .
\end{eqnarray}

Next, we examine the constraints on the gauge potential $A$. Note that 
\begin{eqnarray}
\ell^{-1} A = {\frac{1 }{2}} d \log (1+|\mu|^2)+{\frac{1 }{2}} d \log H + P
du
\end{eqnarray}
for some function $P$. Hence, by making a gauge transformation in $A$,
combined with an appropriately chosen $Spin(3,1)$ transformation generated
by $\Gamma_{+-}$, which together with the $A$-gauge transformation leave 
$1+\mu e^1$ invariant, one can without loss of generality work in a gauge for
which 
\begin{eqnarray}
\ell^{-1} A = \bigg(\frac{1}{2}\frac{\left( \bar{\mu}\partial _{-}\mu +\bar{\mu}
\partial _{-}\mu \right) }{\left( 1+\mu \bar{\mu}\right) } -{\frac{1}{2}}
\omega _{-,+-} \bigg) \mathbf{e}^-
\end{eqnarray}
and moreover, 
\begin{eqnarray}
\omega_{+,+-}=0
\end{eqnarray}
and 
\begin{eqnarray}
-{\frac{1}{2}}\omega _{{1},+-} -{\frac{1}{\sqrt{2}}}\mu {i}\ell ^{-1}=0 \ .
\end{eqnarray}
In this gauge, we then find 
\begin{eqnarray}
d \mathbf{e}^- = -d \log (1+|\mu|^2) \wedge \mathbf{e}^-
\end{eqnarray}
and hence it is most convenient to introduce a local co-ordinate $u$ such
that 
\begin{eqnarray}
\mathbf{e}^- ={\frac{1 }{1+|\mu|^2}} du \ .
\end{eqnarray}
Next, consider the exterior derivative of $\mathbf{e}^1$ restricted to
hypersurfaces $u=const.$. One finds that 
\begin{eqnarray}
{\hat{d}} \mathbf{e}^1 =- {\hat{d}} \log (1+|\mu|^2) \wedge \mathbf{e}^1
\end{eqnarray}
where ${\hat{d}}$ denotes the restriction of the exterior derivative to 
$u=const.$. It follows that one can introduce a complex co-ordinate $z$ such
that 
\begin{eqnarray}  \label{basenull1}
\mathbf{e}^1 = {\frac{1 }{1+|\mu|^2}}(dz+\xi du)
\end{eqnarray}
for $\xi \in \mathbb{C}$. The metric can be simplified further by making use
of the $Spin(3,1)$ gauge transformation generated by $\beta \Gamma_{+1} + 
\bar{\beta} \Gamma_{+ \bar{1}}$, for $\beta \in \mathbb{C}$, which leaves
the Killing spinor $1+\mu e^1$ invariant. This gauge transformation
corresponds to a change of basis of the form 
\begin{eqnarray}
(\mathbf{e}^-)^{\prime }&=& \mathbf{e}^-  \nonumber \\
(\mathbf{e}^+)^{\prime }&=& \mathbf{e}^+ +4 |\beta|^2 \mathbf{e}^- -2 
{\bar{\beta}} \mathbf{e}^1 -2 \beta \mathbf{e}^{\bar{1}}  \nonumber \\
(\mathbf{e}^1)^{\prime }&=& \mathbf{e}^1 -2 \beta \mathbf{e}^- \ .
\end{eqnarray}
By choosing $\beta$ appropriately, one can, without loss of generality, set 
$\xi=0$ in ({\ref{basenull1}}).

So, on introducing a final local co-ordinate $v$ such that the vector field
dual to $\mathbf{e}^-$ is ${\frac{\partial }{\partial v}}$, one finds that
one can write the basis in the $u,v,z, {\bar{z}}$ co-ordinates as 
\begin{eqnarray}  \label{simplerbasis1}
\mathbf{e}^- &=& {\frac{1 }{1+|\mu|^2}} du  \nonumber \\
\mathbf{e}^1 &=& {\frac{1 }{1+|\mu|^2}} dz  \nonumber \\
\mathbf{e}^+ &=& dv + \mathcal{H} du + \mathcal{G} dz + {\bar{\mathcal{G}}} 
d {\bar{z}}
\end{eqnarray}
where $\mathcal{H}$ is a real function, $\mathcal{G}$ is a complex function,
and $\mu$ does not depend on $v$.

Next consider the constraints implied by ({\ref{muder}}). In particular, $\mu
$ depends only on ${\bar{z}}$ and $u$, with 
\begin{eqnarray}
{\frac{\partial \mu }{\partial {\bar{z}}}} = \sqrt{2} i \ell^{-1} \ .
\end{eqnarray}

Hence 
\begin{eqnarray}
\mu = \sqrt{2} i \ell^{-1} {\bar{z}} + h(u)
\end{eqnarray}
where $h$ is a function of $u$. By making a change in co-ordinates of the
form ${\bar{z}}^{\prime }= {\bar{z}} + \psi(u)$ together with an
appropriately chosen  $Spin(3,1)$ transformation generated by $\beta
\Gamma_{+1} + \bar{\beta} \Gamma_{+ \bar{1}}$, one can without loss of
generality take the basis given in ({\ref{simplerbasis1}}) with 
\begin{eqnarray}
\mu = \sqrt{2} i \ell^{-1} {\bar{z}} \ .
\end{eqnarray}
Observe that, in this basis, $\partial_- \mu =0$.

To proceed, consider the conditions $\omega_{-,+1}=\omega_{-,1 \bar{1}}=0$
on the geometry. It is straightforward to show that these imply that 
\begin{eqnarray}
{\mathcal{G}} = -{\frac{2 \ell^{-2} v \bar{z} }{1+2 \ell^{-2} z {\bar{z}}}}+
\phi
\end{eqnarray}
where $\phi(u,z,{\bar{z}})$ is a complex function satisfying 
\begin{eqnarray}
\partial_z \bigg({\bar{\phi} \over 1+2\ell^{-2} z \bar{z}}\bigg)- \partial_{\bar{z}} \bigg({\phi \over 1+2\ell^{-2} z \bar{z}}\bigg) =0 \ .
\end{eqnarray}
Next, noting that 
\begin{eqnarray}
\omega_{-,+-}=- (1+2 \ell^{-2} z {\bar{z}}) 
{\frac{\partial \mathcal{H} }{\partial v}}
\end{eqnarray}
we impose the Bianchi identity $F=dA$ to obtain the conditions
\begin{eqnarray}
{\frac{\partial^2 \mathcal{H} }{\partial v^2}} = {\frac{2 \ell^{-2} }{1+2
\ell^{-2} z {\bar{z}}}}, \qquad {\frac{1 }{2}} {\frac{\partial^2 \mathcal{H} 
}{\partial z \partial v}} = {\frac{\ell^{-2} }{1+2 \ell^{-2} z {\bar{z}}}}
\big( -{\frac{2 \ell^{-2} v \bar{z} }{1+2 \ell^{-2} z {\bar{z}}}}+ \phi\big) 
\ .
\end{eqnarray}
These can be solved to find 
\begin{eqnarray}
\mathcal{H} ={\frac{\ell^{-2} v^2 }{1+2 \ell^{-2} z {\bar{z}}}}+ \Theta_1 v
+ \Theta_2
\end{eqnarray}
where $\Theta_1, \Theta_2$ do not depend on $v$, and 
\begin{eqnarray}
\phi = {\frac{1 }{2}}\ell^2 (1+2 \ell^{-2} z {\bar{z}}) {\frac{\partial
\Theta_1 }{\partial z}} \ .
\end{eqnarray}
One can simplify the solution considerably by making the co-ordinate transformation 
\begin{eqnarray}
v = (1+2 \ell^{-2} z {\bar{z}})  \big( v^{\prime }-{\frac{1 }{2}} \ell^2\Theta_1 \big) \ .
\end{eqnarray}
On dropping the prime on $v^{\prime }$ the solution can then be written as 
\begin{eqnarray}
ds^2 &=& 2 du \bigg[dv + \big(\ell^{-2} v^2 + \Psi \big) du \bigg] +{\frac{2 }{(1+2\ell^{-2} z {\bar{z}})^2}} dz d {\bar{z}}
\end{eqnarray}
with 
\begin{eqnarray}
F =\ell^{-1} dv \wedge du \ .
\end{eqnarray}
The function $\Psi=\Psi(u,z,{\bar{z}})$ appearing in the metric is
constrained to be harmonic on $\mathbb{R}^2$ by the Einstein equations: 
\begin{eqnarray}
{\frac{\partial^2 \Psi }{\partial z \partial {\bar{z}}}} =0 \ .
\end{eqnarray}
Observe that the gauge field equations $d \star F=0$ hold with no further
constraints.

\section{Solutions with $\protect\epsilon =1+ e^{i \protect\alpha} e_{2}$}

On evaluating the equations in Appendix A with $\lambda=1$, $\mu^1=0$, 
$\mu^2 = e^{i \alpha}$, one obtains the components of the gauge field
strength as:

\begin{eqnarray}
\label{fluxc1}
F_{+-} &=&\sqrt{2} \big( \sin \alpha \omega
_{+,+-}-\partial _{+}\alpha \cos \alpha \big) - \ell^{-1} \nonumber \\
F_{1{\bar{1}}} &=&i\sqrt{2} \big( \cos \alpha \omega _{+,+-}+\sin \alpha
\partial _{+}\alpha \big)  \nonumber \\
F_{-1} &=&\frac{i}{\sqrt{2}}e^{i\alpha }\omega _{-,-1}  \nonumber \\
F_{+1} &=&\frac{i}{\sqrt{2}}e^{-i\alpha }\omega _{+,+1} \ .
\end{eqnarray}
The components of the gauge potential are given by: 
\begin{equation}
\label{fluxc2}
\ell ^{-1}A_{-}=-{\frac{1}{2}}\omega _{-,+-}, \ \ \ \ell ^{-1}A_{1}= 
\frac{1}{2}\left( i\partial _{1}\alpha -\omega _{1,1{\bar{1}}}\right) , 
\ \ \ \ell ^{-1}A_{+}=\frac{1}{2}\omega _{+,+-} \ .
\end{equation}
The geometric constraints are given by 
\begin{eqnarray}
\omega _{+,+-}+\omega _{-,+-} &=&\sqrt{2}\ell ^{-1}\sin \alpha  \nonumber \\
\partial _{-}\alpha -\partial _{+}\alpha &=&\sqrt{2}\ell ^{-1}\cos \alpha 
\nonumber \\
\omega _{{\bar{1}},1{\bar{1}}} &=&2i\partial _{\bar{1}}\alpha +
\omega _{+,+\bar{1}}=\omega _{-,-\bar{1}}  \nonumber \\
\omega _{{\bar{1}},+1} &=&-\omega _{+,+-}-i\partial _{+}\alpha -i\sqrt{2}
e^{i\alpha }\ell ^{-1}  \nonumber \\
\omega _{{\bar{1}},-1} &=&-\omega _{+,+-}+i\partial _{+}\alpha  \nonumber \\
\omega _{1,+-} &=&-i\partial _{1}\alpha , \ \ 
\omega _{+,1{\bar{1}}}=2i\partial _{+}\alpha  \nonumber \\
\omega _{-,1{\bar{1}}} &=&\omega _{+,-1}=\omega _{-,+1}=\omega
_{1,+1}=\omega _{1,-1}=0 \ .
\end{eqnarray}
Thus we can write \bigskip 
\begin{eqnarray}
\label{extder1}
d\mathbf{e}^{1} &=&\left( -\omega _{+,+-}+i\sqrt{2}e^{-i\alpha }\ell
^{-1}-i\partial _{+}\alpha \right) \mathbf{e}^{1}\wedge 
\mathbf{e}^{+}
\nonumber  \\
&-&\left( \omega _{+,+-}+i\partial _{+}\alpha \right) \mathbf{e}^{1}\wedge 
\mathbf{e}^{-}
-\left( 2i\partial _{\bar{1}}\alpha + \omega _{+,+\bar{1}}\right) 
\mathbf{e}^{1}\wedge \mathbf{e}^{\bar{1}}  \nonumber \\
d\mathbf{e}^{+} &=&\omega _{-,-+}\mathbf{e}^{+}\wedge \mathbf{e}^{-}+\omega
_{-,-1}\mathbf{e}^{1}\wedge \mathbf{e}^{-}-i\partial _{1}\alpha 
\mathbf{e}^{1}\wedge \mathbf{e}^{+}
\nonumber \\
&+&\omega _{-,-\bar{1}}\mathbf{e}^{\bar{1}}\wedge 
\mathbf{e}^{-}+i\partial _{\bar{1}}\alpha \mathbf{e}^{\bar{1}}\wedge 
\mathbf{e}^{+}  +2i\partial _{+}\alpha \mathbf{e}^{1}\wedge \mathbf{e}^{\bar{1}}  \nonumber \\
d\mathbf{e}^{-} &=&-\omega _{++-}\mathbf{e}^{+}\wedge \mathbf{e}^{-}+i\left(
\partial _{1}\alpha \mathbf{e}^{1}-\partial _{\bar{1}}\alpha 
\mathbf{e}^{\bar{1}}\right) \wedge \mathbf{e}^{-}
\nonumber \\
&+&\left( \omega _{+,+1} 
\mathbf{e}^{1}+\omega _{+,+\bar{1}}\mathbf{e}^{\bar{1}}\right) \wedge \mathbf{e}^{+} 
-2i\left( \partial _{+}\alpha +\sqrt{2}\cos \alpha \ell ^{-1}\right) 
\mathbf{e}^{1}\wedge \mathbf{e}^{\bar{1}} \ .
\end{eqnarray}

\subsection{Solutions with $\cos \protect\alpha \neq 0$}

For these solutions, it is convenient to define the 1-form 
\begin{eqnarray}
V = {\frac{1 }{\cos \alpha}} (\mathbf{e}^+ - \mathbf{e}^-)
\end{eqnarray}
and introduce a local co-ordinate $t$ such that $V={\frac{\partial }{\partial t}}$.

It is straightforward to see that the supersymmetry constraints imply that 
\begin{eqnarray}
{\frac{\partial \alpha }{\partial t}} = \sqrt{2} \ell^{-1}
\end{eqnarray}
and furthermore 
\begin{eqnarray}
\mathcal{L}_V \mathbf{e}^1 &=& {\frac{\sqrt{2} i \ell^{-1} 
e^{-i \alpha} }{\cos \alpha}} \mathbf{e}^1  \nonumber \\
\mathcal{L}_V (\mathbf{e}^+ + \mathbf{e}^-) &=& \sqrt{2} \ell^{-1} \tan
\alpha (\mathbf{e}^+ + \mathbf{e}^-) \ .
\end{eqnarray}
These constraints imply that one can write 
\begin{eqnarray}
\mathbf{e}^1 &=& \big(1+i \tan \alpha \big) {\hat{\mathbf{e}}}^1  \nonumber \\
\mathbf{e}^+ + \mathbf{e}^- &=& {\frac{\sqrt{2} }{\cos \alpha}} \mathbf{e}^2
\end{eqnarray}
where 
\begin{eqnarray}  \label{liederv1}
\mathcal{L}_V {\hat{\mathbf{e}}}^1 =0, \qquad \mathcal{L}_V \mathbf{e}^2 =0
\ .
\end{eqnarray}

Note, furthermore, that 
\begin{eqnarray}
\label{based1}
d {\hat{\mathbf{e}}}^1 &=& \bigg( \sqrt{2} \sec \alpha \omega_{+,+-}
-{1 \over \sqrt{2}} {\sin \alpha \over \cos^2 \alpha}(\partial_+ \alpha + \partial_- \alpha)
\nonumber \\
&-& \ell^{-1} \tan \alpha - 2i \ell^{-1} \bigg) 
\mathbf{e}^2 \wedge {\hat{\mathbf{e}}}^1  \nonumber \\
&+& \sec^2 \alpha (-i \partial_{\bar{1}} \alpha - \cos \alpha e^{-i \alpha}
\omega_{+,+ \bar{1}}) {\hat{\mathbf{e}}}^1 \wedge
 {\hat{\mathbf{e}}}^{\bar{1}} \ .
\end{eqnarray}
and
\begin{eqnarray}
\label{based2}
d {\bf{e}}^2 &=& 
{1 \over 2} \sec^2 \alpha \bigg( (e^{2i \alpha} \omega_{+,+1}+\omega_{-,-1}){\hat{{\bf{e}}}}^1
+(e^{-2i\alpha} \omega_{+,+\bar{1}}+\omega_{-,-\bar{1}}) {\hat{{\bf{e}}}}^{\bar{1}} \bigg) \wedge {\bf{e}}^2
\nonumber \\
&-&2i\ell^{-1} {\hat{{\bf{e}}}}^1 \wedge {\hat{{\bf{e}}}}^{\bar{1}}
\end{eqnarray}

Note then that the metric can be written as
\begin{eqnarray}
ds^2 = -{1 \over 2} ({\bf{e}}^+-{\bf{e}}^-)^2 + \sec^2 \alpha \ ds^2_{GT}
\end{eqnarray}
where
\begin{eqnarray}
ds^2_{GT} = ({\bf{e}}^2)^2 +2 {\hat{{\bf{e}}}}^1 {\hat{{\bf{e}}}}^{\bar{1}}
\end{eqnarray}
The metric on the manifold $GT$ does not depend on $t$, and
({\ref{based1}}) and ({\ref{based2}}) imply that $GT$
admits a $t$-independent basis ${\bf{E}}^i$ for $i=1,2,3$ satisfying
\begin{eqnarray}
\label{gaudtod}
d {\bf{E}}^i = {\cal{B}} \wedge {\bf{E}}^i -\ell^{-1} \epsilon^{ijk} {\bf{E}}^j \wedge {\bf{E}}^k
\end{eqnarray}
where
\begin{eqnarray}
{\cal{B}}&=& {1 \over 2}\sec^2 \alpha \bigg((\omega_{-,-1}+e^{2i\alpha} \omega_{+,+1}) {\hat{{\bf{e}}}}^1
+ (\omega_{-,-\bar{1}}+e^{-2i\alpha} \omega_{+,+\bar{1}}) {\hat{{\bf{e}}}}^{\bar{1}} \bigg)
\nonumber \\
&+& \bigg(\sqrt{2} \sec \alpha \omega_{+,+-} -{1 \over \sqrt{2}}{\sin \alpha \over
\cos^2 \alpha} (\partial_+ \alpha + \partial_- \alpha)-\ell^{-1} \tan \alpha \bigg) {\bf{e}}^2 \ .
\end{eqnarray}
Note in particular that ({\ref{gaudtod}}) implies that  ${\cal{B}}$ must be independent of $t$, and
furthermore, must satisfy
\begin{eqnarray}
\label{intcond1}
d {\cal{B}} = 2 \ell^{-1} \star_3 {\cal{B}}
\end{eqnarray}
where $\star_3$ denotes the Hodge dual on $GT$ (in our conventions, the volume form on $GT$ is
$i {\hat{{\bf{e}}}}^1 \wedge {\hat{{\bf{e}}}}^{\bar{1}} \wedge {\bf{e}}^2$). In turn,
({\ref{intcond1}}) implies that 
\begin{eqnarray}
\label{intcond2}
d \star_3 {\cal{B}} =0 \ .
\end{eqnarray}
The condition ({\ref{gaudtod}}) imples that $GT$ admits a Gauduchon-Tod structure. Such structures
arise in the context of 4-dimensional hyper-K\"ahler with torson manifolds
which admit a tri-holomorphic isometry, and have been analysed in \cite{gtod} and \cite{mactod}.

To proceed further, we next consider the constraints which ({\ref{extder1}}) impose on ${\bf{e}}^+-{\bf{e}}^-$.
It will be convenient to write
\begin{eqnarray}
{\bf{e}}^+ -{\bf{e}}^- = -2 \sec \alpha (dt+\Omega)
\end{eqnarray}
and to set 
\begin{eqnarray}
\alpha = \sqrt{2} \ell^{-1} t + \Phi
\end{eqnarray}
where $\Omega$ is a $t$-dependent 1-form on $GT$ and $\Phi$ is a $t$-independent function.
Then ({\ref{extder1}}) implies that
\begin{eqnarray}
{\cal{L}}_V \Omega +2 \sqrt{2} \ell^{-1} \tan \alpha \ \Omega - {\cal{B}} - 2 \tan \alpha \ d \Phi =0 \ .
\end{eqnarray}
This condition can be integrated up, and on changing co-ordinates from $t$ to $\alpha$, one obtains
\begin{eqnarray}
{\bf{e}}^+ - {\bf{e}}^- = -\sqrt{2} \ell \sec \alpha d \alpha -\sqrt{2} \ell \sin \alpha \ {\cal{B}} - 2 \cos \alpha \ \psi
\end{eqnarray}
where $\psi$ is an $\alpha$-independent 1-form on $GT$.
Note that $\psi$ is defined in terms of the basis ${\bf{e}}^1, {\hat{{\bf{e}}}}^{\bar{1}}, {\bf{e}}^2$ as
\begin{eqnarray}
\psi &=& {\ell \over 2 \sqrt{2} \cos^2 \alpha} \bigg(i (-\omega_{-,-1}+e^{2i\alpha} \omega_{+,+1})
{\hat{{\bf{e}}}}^1 -i (-\omega_{-,- \bar{1}}+e^{-2i\alpha} \omega_{+,+\bar{1}}) {\hat{{\bf{e}}}}^{\bar{1}} \bigg)
\nonumber \\
&+&{1 \over 2}\bigg(-\ell \sec \alpha (\partial_+\alpha+\partial_-\alpha)-\sqrt{2} \ell
{\sin \alpha \over \cos^2 \alpha}(\sqrt{2} \omega_{+,+-}-\ell^{-1} \sin \alpha) \bigg) {\bf{e}}^2 .
\end{eqnarray}
The remaining content of ({\ref{extder1}}) imposes an additional condition on $\psi$:
\begin{eqnarray}
\label{psicon1}
d \psi + {\cal{B}} \wedge \psi -2 \ell^{-1} \star_3 \psi =0 \ .
\end{eqnarray}
It remains to consider the constraints on the fluxes. Note first that
({\ref{fluxc2}}) implies that
\begin{eqnarray}
\label{fluxpotx1}
A = -{\ell \over 2} \tan \alpha \ d \alpha + {\ell \over 2} \cos 2 \alpha \ {\cal{B}} -{1 \over \sqrt{2}} \sin 2 \alpha \ \psi \ .
\end{eqnarray}
It is straightforward to show that on applying the exterior derivative to ({\ref{fluxpotx1}}),
one obtains the components of the field strength given in ({\ref{fluxc1}}), with no further constraint.
In order to evaluate the gauge field equations, observe that the above conditions imply that the
Hodge dual of $F$ is given by
\begin{eqnarray}
\star F &=& -\sqrt{2} \ell d \alpha \wedge \bigg({1 \over \sqrt{2}} \cos 2 \alpha \ {\cal{B}} - \ell^{-1} \sin 2 \alpha \ \psi \bigg)
\nonumber \\ 
&+& \sqrt{2} \cos^2 \alpha \ {\cal{B}} \wedge \psi - \sin 2 \alpha \star_3 {\cal{B}} - \sqrt{2} \ell^{-1} \cos 2 \alpha \star_3 \psi \ .
\end{eqnarray}
The conditions obtained previously imply that the RHS of this expression is closed with no additional
constraint, hence the gauge equations are satisfied.

In order to examine the Einstein equations we follow the reasoning presented (in the context of
solutions of the anti-de-Sitter minimal gauged supergravity) in Appendix E of
\cite{klemm2003}. In particular, the integrability conditions of the Killing spinor equation associated with
a pseudo-supersymmetric solution for which the Maxwell field strength $F$ satisfies the Bianchi identity
and gauge field equations imply that
\begin{eqnarray}
\label{einint}
E_{\mu \nu} \Gamma^\nu \epsilon =0
\end{eqnarray}
where
\begin{eqnarray}
E_{\mu \nu} = R_{\mu \nu} -3 \ell^{-2} g_{\mu \nu} -2 F_{\mu \rho} F_{\nu}{}^\rho +{1 \over 2} 
F_{\alpha \beta}F^{\alpha \beta} g_{\mu \nu} \ .
\end{eqnarray}
Evaluating ({\ref{einint}}) acting on the Killing spinor $\epsilon=1+e^{i \alpha} e_2$, one finds that
all components of $E_{\mu \nu}$ are constrained to vanish, i.e. the Einstein equations hold
automatically.

To summarize, the solutions with Killing spinor $1+ e^{i \protect\alpha} e_{2}$ and $\cos \alpha \neq 0$ have
metric 
\begin{eqnarray}
ds^2 = - \big(\ell \sec \alpha \ d \alpha + \ell \sin \alpha \ {\cal{B}} + \sqrt{2} \cos \alpha \ \psi \big)^2
+ \sec^2 \alpha ds^2_{GT}
\end{eqnarray}
where $ds^2_{GT}$ is an $\alpha$-independent metric on a 3-dimensional manifold which has  a Gauduchon-Tod structure.
The 3-manifold $GT$ admits a $\alpha$-independent basis ${\bf{E}}^i$ and an $\alpha$-independent 1-form
${\cal{B}}$ satisfying ({\ref{gaudtod}}) (with associated integrability conditions ({\ref{intcond1}}) and ({\ref{intcond2}})).
$GT$ also admits an $\alpha$-independent 1-form $\psi$ satisfying ({\ref{psicon1}}). The flux is then given by
\begin{eqnarray}
F = d \bigg(  {\ell \over 2} \cos 2 \alpha \ {\cal{B}} -{1 \over \sqrt{2}} \sin 2 \alpha \ \psi \bigg) \ .
\end{eqnarray}
Finally, we remark that on making the co-ordinate transformation $t'=\ell \tan \alpha$, one finds that the
metric can be written in the form originally obtained in \cite{meessen}.

\subsection{Solutions with $\cos \protect\alpha =0$}

Suppose that $\sin \alpha = \pm 1,$ then

\begin{eqnarray}  \label{gaugefs1}
F &=& \big(\pm \sqrt{2} \omega_{+,+-}- \ell^{-1}\big) \mathbf{e}^+ \wedge 
\mathbf{e}^- 
\nonumber \\
&\pm& {\frac{1 }{\sqrt{2}}} \big(\mathbf{e}^+-\mathbf{e}^-)
\wedge \big(\omega_{+,+1}\mathbf{e}^1 
+\omega_{+,+\bar{1}} \mathbf{e}^{\bar{1}} \big)
\end{eqnarray}
and 
\begin{eqnarray}  \label{gaugepot1}
\ell^{-1} A = {\frac{1 }{2}} \omega_{+,+-} \mathbf{e}^+ -{\frac{1 }{2}}
\omega_{-,+-} \mathbf{e}^- +{\frac{1 }{2}} \omega_{+,+1} \mathbf{e}^1 
+{\frac{1 }{2}}\omega_{+,+\bar{1}} \mathbf{e}^{\bar{1}}
\end{eqnarray}
and

\begin{eqnarray}
\omega _{+,+-}+\omega _{-,+-} &=& \pm \sqrt{2}\ell ^{-1}  \nonumber \\
\hskip15mm \omega _{{\bar{1}},1{\bar{1}}} &=&\omega _{+,+\bar{1}}=\omega _{-,-\bar{1}} 
\nonumber \\
\omega _{{\bar{1}},+1}+\omega _{+,+-} &=&\pm \sqrt{2}\ell ^{-1}  \nonumber \\
\hskip15mm \omega _{{\bar{1}},-1} &=&-\omega _{+,+-}  \nonumber \\
\hskip15mm \omega _{1,+-} &=&\omega _{+,1{\bar{1}}}=\omega _{-,1{\bar{1}}}=\omega
_{+,-1}=\omega _{-,+1}=\omega _{1,+1}=\omega _{1,-1}=0 \ .
\end{eqnarray}
It follows that 
\begin{eqnarray}  \label{extalg}
d \mathbf{e}^+ &=& -\omega_{-,+-} \mathbf{e}^+ \wedge \mathbf{e}^- + \big(
\omega_{+,+1}\mathbf{e}^1 + \omega_{+,+\bar{1}} \mathbf{e}^{\bar{1}} \big) 
\wedge \mathbf{e}^-  \nonumber \\
d \mathbf{e}^- &=& -\omega_{+,+-} \mathbf{e}^+ \wedge \mathbf{e}^- + \big(
\omega_{+,+1}\mathbf{e}^1 + \omega_{+,+\bar{1}} \mathbf{e}^{\bar{1}} \big) 
\wedge \mathbf{e}^+  \nonumber \\
d \mathbf{e}^1 &=& \bigg[ \big(\omega_{+,+-}\mp \sqrt{2} \ell^{-1} \big) 
\mathbf{e}^+ + \omega_{+,+-} \mathbf{e}^- 
+\omega_{+,+ \bar{1}}\mathbf{e}^{\bar{1}} \bigg] \wedge \mathbf{e}^1 \ .
\end{eqnarray}
To proceed, note that ({\ref{extalg}}) implies that 
\begin{eqnarray}
(\mathbf{e}^+ +\mathbf{e}^-) \wedge \mathrm{d}
 (\mathbf{e}^++\mathbf{e}^-)=0, \qquad (\mathbf{e}^+ -\mathbf{e}^-) \wedge \mathrm{d} 
(\mathbf{e}^+-\mathbf{e}^-)=0 \ .
\end{eqnarray}
Hence, there exist real functions $H, B,z,t$ such that 
\begin{eqnarray}
\mathbf{e}^+ = {\frac{1 }{\sqrt{2}}} \big(H dz - B dt), \qquad \mathbf{e}^-
= {\frac{1 }{\sqrt{2}}} (H dz+B dt) \ .
\end{eqnarray}
Next, note that ({\ref{gaugefs1}}) and ({\ref{extalg}}) imply that 
\begin{eqnarray}
F = \pm {\frac{1 }{\sqrt{2}}} d (\mathbf{e}^+ - \mathbf{e}^-) \ .
\end{eqnarray}
On comparing this expression with ({\ref{gaugepot1}}), one finds that there
exists a function $C$ such that 
\begin{eqnarray}
{\frac{1 }{\sqrt{2}}} (\mathbf{e}^+ - \mathbf{e}^-) &=& {\frac{\ell }{2}} 
\big( \omega_{+,+-} \mathbf{e}^+ -\omega_{-,+-}\mathbf{e}^- +\omega_{+,+1}
\mathbf{e}^1+\omega_{+,+\bar{1}}\mathbf{e}^{\bar{1}} \big)
\nonumber \\
&-&{\frac{\ell }{2}} d \log C \ .
\end{eqnarray}
Substituting this expression back into ({\ref{extalg}}) one finds 
\begin{eqnarray}
d \mathbf{e}^1 = d \log C \wedge \mathbf{e}^1
\end{eqnarray}
and so there exist real functions $C$, $x$, $y$ such that 
\begin{eqnarray}
\mathbf{e}^1 = {\frac{1 }{\sqrt{2}}} C (dx+idy) \ .
\end{eqnarray}
It is then straightforward to show that ({\ref{extalg}}) implies that 
\begin{eqnarray}
H = C f_1(z), \qquad B = C^{-1} f_2(t)
\end{eqnarray}
where $f_1$ and $f_2$ are arbitrary functions of $z$, $t$. By making
appropriate $z$, $t$ co-ordinate transformations, one can without loss of
generality take $f_1=f_2=1$. Furthermore, ({\ref{extalg}}) implies that 
\begin{eqnarray}
{\frac{\partial C }{\partial t}} = \pm \ell^{-1}
\end{eqnarray}
so that 
\begin{eqnarray}
C = \pm \ell^{-1} t + V
\end{eqnarray}
for $V=V(x,y,z)$. The metric and gauge field strength are then given by 
\begin{eqnarray}  \label{cosmp}
ds^2 =- {\frac{1 }{(V \pm \ell^{-1} t)^2}} dt^2 + (V \pm \ell^{-1} t)^2 
\big(dx^2+dy^2+dz^2 \big)
\end{eqnarray}
and 
\begin{eqnarray}
F = \mp d \bigg( {\frac{1 }{V \pm \ell^{-1} t}} dt \bigg) \ .
\end{eqnarray}
Finally, we impose the gauge field equations $d \star F=0$, which imply that 
$V$ is harmonic on $\mathbb{R}^3$: 
\begin{eqnarray}
\bigg( {\frac{\partial^2 }{\partial x^2}} + 
{\frac{\partial^2 }{\partial y^2}}+{\frac{\partial^2 }{\partial z^2}} \bigg) V =0 \ ,
\end{eqnarray}
and we remark that, from the reasoning used in the previous sub-section, this 
condition is sufficient to ensure that the Einstein equations hold automatically.

This solution is the cosmological Majumdar-Papapetrou black hole solution
found in \cite{kastortraschen}. Observe that on taking the limit $\ell
\rightarrow \infty $ one recovers the standard Majumdar-Papapetrou solution.
The cosmological solution ({\ref{cosmp}}) is obtained by shifting the
harmonic function by a term linear in $t$; this method of obtaining
solutions in de Sitter supergravity has also been investigated in \cite{hkt,
berncvetic}

\section{Conclusions}

Using spinorial geometry techniques, all pseudo-supersymmetric solutions of minimal 
de Sitter $N=2$, $D=4$ supergravity have been classified. There are three classes of
solutions:

\begin{itemize}
\item[(i)] The first class of solution has metric and field strength 
\begin{eqnarray}
\label{grav}
ds^2 &=& 2 du \bigg[dv + \big(\ell^{-2} v^2 + \Psi \big) du \bigg] +{\frac{2 }{(1+2\ell^{-2} z {\bar{z}})^2}} dz d {\bar{z}}
\end{eqnarray}
with 
\begin{eqnarray}
F =\ell^{-1} dv \wedge du 
\end{eqnarray}
where $\Psi=\Psi(u,z,{\bar{z}})$ satisfies 
\begin{eqnarray}
{\frac{\partial^2 \Psi }{\partial z \partial {\bar{z}}}} =0 \ .
\end{eqnarray}

\item[(ii)]  The second class of solution has metric
\begin{eqnarray}
ds^2 = - \big(\ell \sec \alpha \ d \alpha + \ell \sin \alpha \ {\cal{B}} + \sqrt{2} \cos \alpha \ \psi \big)^2
+ \sec^2 \alpha ds^2_{GT}
\end{eqnarray}
where $ds^2_{GT}$ is an $\alpha$-independent metric on a 3-dimensional manifold which has  a Gauduchon-Tod structure.
The 3-manifold $GT$ admits a $\alpha$-independent basis ${\bf{E}}^i$ and an $\alpha$-independent 1-form
${\cal{B}}$ satisfying
\begin{eqnarray}
d {\bf{E}}^i = {\cal{B}} \wedge {\bf{E}}^i -\ell^{-1} \epsilon^{ijk} {\bf{E}}^j \wedge {\bf{E}}^k
\end{eqnarray}
together with an $\alpha$-independent 1-form $\psi$ satisfying 
\begin{eqnarray}
d \psi + {\cal{B}} \wedge \psi -2 \ell^{-1} \star_3 \psi =0 \ .
\end{eqnarray}
The gauge field strength is
\begin{eqnarray}
F = d \bigg(  {\ell \over 2} \cos 2 \alpha \ {\cal{B}} -{1 \over \sqrt{2}} \sin 2 \alpha \ \psi \bigg) \ .
\end{eqnarray}

\item[(iii)] The third class of solution consists of the cosmological
Majumdar-Papapetrou black hole solution found in \cite{kastortraschen} with 
\begin{eqnarray}  \label{cosmp2}
ds^2 =- {\frac{1 }{(V \pm \ell^{-1} t)^2}} dt^2 + 
(V \pm \ell^{-1} t)^2 \big(dx^2+dy^2+dz^2 \big)
\end{eqnarray}
and 
\begin{eqnarray}
F = \mp d \bigg( {\frac{1 }{V \pm \ell^{-1} t}} dt \bigg) \ 
\end{eqnarray}
where $V=V(x,y,z)$ satisfies 
\begin{eqnarray}
\bigg( {\frac{\partial^2 }{\partial x^2}} + 
{\frac{\partial^2 }{\partial y^2}}+{\frac{\partial^2 }{\partial z^2}} \bigg) V =0 \ .
\end{eqnarray}
\end{itemize}

We remark that the solution given in $(i)$ for the special case $\Psi=0$ was found in \cite{meessen};
where it is noted that the spacetime is the Nariai solution \cite{nariai};
the solution ({\ref{grav}}) corresponds to gravitational waves propagating in this background \cite{podol}.

\appendix

\section{The Linear System}

In this appendix we present the decomposition of the Killing spinor equation
acting on the spinor $\epsilon =\lambda 1+\mu ^{i}e_{i}$; we obtain the
following constraints:

{\small 
\begin{equation*}
\partial _{+}\lambda +\lambda \left( -{\frac{1}{2}}\omega _{+,+-} 
-{\frac{1}{2}}\omega _{+,1{\bar{1}}}-\ell ^{-1}A_{+}\right) -\frac{i}{\sqrt{2}}\mu
^{2}\left( F_{+-}+F_{1{\bar{1}}}+\ell ^{-1}\right) =0
\end{equation*}

\begin{equation*}
\partial _{+}\mu ^{1}+\mu ^{1}\left( -{\frac{1}{2}}\omega _{+,+-} 
+{\frac{1}{2}}\omega _{+,1{\bar{1}}}-\ell ^{-1}A_{+}\right) -\omega _{+,-1}\mu ^{2}=0
\end{equation*}

\begin{equation*}
\partial _{+}\mu ^{2}+\omega _{+,+{\bar{1}}}\mu ^{1}+\mu ^{2} 
\left({\frac{1}{2}}\omega _{+,+-}-{\frac{1}{2}}\omega _{+,1{\bar{1}}}
-\ell^{-1}A_{+}\right) =0
\end{equation*}

\begin{equation*}
\omega _{+,+1}\lambda +\sqrt{2}iF_{+1}\mu ^{2}=0
\end{equation*}

\begin{equation*}
\partial _{-}\lambda +\lambda \left( -{\frac{1}{2}}\omega _{-,+-} 
-{\frac{1}{2}}\omega _{-,1{\bar{1}}}-\ell ^{-1}A_{-}\right)
 -\sqrt{2} iF_{-{\bar{1}}}\mu^{1}=0
\end{equation*}

\begin{equation*}
\partial _{-}\mu ^{1}-\sqrt{2}iF_{-1}\lambda +\mu ^{1}\left( 
-{\frac{1}{2}}\omega _{-,+-}+{\frac{1}{2}}\omega _{-,1{\bar{1}}}-\ell ^{-1}A_{-}\right)
-\omega _{-,-1}\mu ^{2}=0
\end{equation*}

\begin{eqnarray*}
\partial _{-}\mu ^{2}+\frac{i}{\sqrt{2}}\lambda \left( F_{+-} 
-F_{1{\bar{1}}}-\ell ^{-1}\right) +\omega _{-,+{\bar{1}}}\mu ^{1}
+\mu ^{2} 
\left({\frac{1}{2}}\omega _{-,+-}-{\frac{1}{2}}\omega _{-,1{\bar{1}}}-\ell ^{-1}A_{-}\right)
=0
\end{eqnarray*}

\begin{equation*}
-\omega _{-,+1}\lambda -\frac{i}{\sqrt{2}}\mu ^{1}\left( F_{+-} 
+F_{1{\bar{1}}}-\ell ^{-1}\right) =0
\end{equation*}

\begin{equation*}
\partial _{1}\lambda +\lambda \left( -{\frac{1}{2}}\omega _{1,+-} 
-{\frac{1}{2}}\omega _{1,1{\bar{1}}}-\ell ^{-1}A_{1}\right) 
-{\frac{i}{\sqrt{2}}}\mu^{1}\left( F_{+-}+F_{1{\bar{1}}}+\ell ^{-1}\right) =0
\end{equation*}

\begin{equation*}
\partial _{1}\mu ^{1}+\mu ^{1}\left( -{\frac{1}{2}}\omega _{1,+-} 
+{\frac{1}{2}}\omega _{1,1{\bar{1}}}-\ell ^{-1}A_{1}\right) -\omega _{1,-1}\mu ^{2}=0
\end{equation*}

\begin{equation*}
\partial _{1}\mu ^{2}+\omega _{1,+{\bar{1}}}\mu ^{1}+\mu ^{2} 
\left( {\frac{1}{2}}\omega _{1,+-}-{\frac{1}{2}}\omega _{1,1{\bar{1}}}
-\ell^{-1}A_{1}\right) =0
\end{equation*}

\begin{equation*}
\omega _{1,+1}\lambda +\sqrt{2}iF_{+1}\mu ^{1}=0
\end{equation*}

\begin{equation*}
\partial _{\bar{1}}\lambda +\lambda \left( -{\frac{1}{2}} 
\omega _{{\bar{1}},+-}-{\frac{1}{2}}\omega _{{\bar{1}},1{\bar{1}}}-\ell ^{-1} 
A_{{\bar{1}}}\right) +\sqrt{2}iF_{-{\bar{1}}}\mu ^{2}=0
\end{equation*}

\begin{equation*}
\partial _{\bar{1}}\mu ^{1}+\lambda \frac{i}{\sqrt{2}}\left( -F_{+-}
 +F_{1{\bar{1}}}-\ell ^{-1}\right) +\mu ^{1}\left( -{\frac{1}{2}} 
\omega _{{\bar{1}},+-}+{\frac{1}{2}}\omega _{{\bar{1}},1{\bar{1}}}-\ell ^{-1} 
A_{\bar{1}}\right) -\omega _{{\bar{1}},-1}\mu ^{2}=0
\end{equation*}

\begin{equation*}
\partial _{\bar{1}}\mu ^{2}+\sqrt{2}iF_{+{\bar{1}}}\lambda +
\omega _{{\bar{1}},+{\bar{1}}}\mu ^{1}+\mu ^{2}\left( {\frac{1}{2}} \omega _{{\bar{1}},+-}
-{\frac{1}{2}}\omega _{{\bar{1}},1{\bar{1}}} -\ell ^{-1}A_{\bar{1}}\right) =0
\end{equation*}

\begin{equation}  \label{ksedecomp}
-\omega _{{\bar{1}},+1}\lambda +\frac{i}{\sqrt{2}}\mu ^{2}\left( F_{+-} 
+F_{1{\bar{1}}}-\ell ^{-1}\right) =0 \ .
\end{equation}
}

\bigskip

\textbf{Acknowledgements:} J. G. ~ is supported by the EPSRC grant,
EP/F069774/1.

\end{document}